\documentclass[12pt]{iopart}
\expandafter\let\csname equation*\endcsname\relax
\expandafter\let\csname endequation*\endcsname\relax
\usepackage{amsmath}
\usepackage{verbatim}
\usepackage{graphicx}

 \newcommand{\ket}[1]{\left|#1\right\rangle} 

\begin{document}


\title{Proximity effects in cold atom artificial graphene}

\author{ Tobias Gra\ss$^{1,2}$, Ravindra W. Chhajlany$^3$, Leticia Tarruell$^2$, Vittorio Pellegrini$^{4,5}$, Maciej Lewenstein$^{2,6,7}$}

\address{$^1$Joint Quantum Institute, University of Maryland, College Park, MD 20742, U.S.A.}
\address{$^2$ICFO-Institut de Ci\`encies Fot\`oniques, The Barcelona Institute of Science and Technology, 08860 Castelldefels (Barcelona), Spain}
\address{$^3$ Faculty of Physics, Adam Mickiewicz University, Umultowska 85, 61-614 Pozna{\'n}, Poland}
\address{$^4$National Enterprise for Nanoscience and Nanotechnology (NEST), Istituto Nanoscienze-Consiglio Nazionale delle Ricerche and Scuola Normale Superiore, I-56126 Pisa, Italy}
\address{$^5$Istituto Italiano di Tecnologia, Graphene labs, Via Morego 30, I-16163 Genova, Italy}
\address{$^6$ICREA-Instituci\'o Catalana de Recerca i
Estudis Avan\c{c}ats, Llu\'is Companys 23, 08010 Barcelona, Spain}
\address{$^7$ Kavli Institute for Theoretical Physics China, Chinese Academy of Sciences, Beijing 100190, China}

\begin{abstract}
Cold atoms in an optical lattice with brick-wall geometry have
been used to mimic graphene, a two-dimensional material with
characteristic Dirac excitations. Here we propose to bring such
artificial graphene into the proximity of a second atomic layer
with a square lattice geometry. For non-interacting fermions, we
find that such bilayer system undergoes a phase transition from a
graphene-like semi-metal phase, characterized by a band structure
with Dirac points, to a gapped band insulator phase. In the
presence of attractive interactions between fermions with
pseudospin-1/2 degree of freedom, a competition between semi-metal
and superfluid behavior is found at the mean-field level.
Using the quantum Monte Carlo method, we also investigate the case of strong repulsive
interactions. In the Mott phase, each layer exhibits a different
amount of long-range magnetic order. Upon coupling both layers, a
valence-bond crystal is formed at a critical coupling strength.
Finally, we discuss how these bilayer systems could be realized in
existing cold atom experiments.
\end{abstract}

\maketitle

\section{Introduction}
The outstanding properties of graphene have boosted the interest
in this two-dimensional, carbon-made material. Technology
developments are linked mostly to its mechanical performances,
large specific surface area and conductivity properties
\cite{roadmap-graphene,castro_neto}. On the fundamental side, the
most intriguing properties emerge from the peculiar material's
band structure originating from the hexagonal geometry of the
lattice, equivalent to two triangular Bravais lattices shifted
relative to each other: each band is split into two separated
ones. These bands touch each other at two distinct points within
the first Brillouin zone, the so-called Dirac points. In the
vicinity of the Dirac points, the dispersion is linear, and
low-energy excitations behave like relativistic particles.
Recently, there have been proposals to modify the band structure
by bringing graphene on top of another graphene layer
\cite{bilayer-graphene}, or close to other substrates
\cite{ortix2012}. Furthermore, the emergence of additional Dirac
points has been observed in graphene on hexagonal boron nitride
\cite{yankowitz2012}. The proximity of graphene to a normal
two-dimensional electron gas was studied in Refs.
\cite{principi12,gamucci14,aliaj16}.

A different approach to studying properties of the graphene band
structure is the use of artificial graphene \cite{maciek-ag}, that
is, of systems that are designed to mimic graphene. Artificial
graphene offers several control parameters allowing to tune the
system. The variety of artificial graphene systems reaches from
solid-state devices such as semiconductor structures with
engineered nanopatterns \cite{park2009,pellegrini,wang16}, or
molecules arranged on metal surfaces \cite{gomesKK}, to optical
systems such as hexagonal photonic crystals \cite{beenakker07,
RechtsmanPRL2013}, microwave fields in metamaterial structures
\cite{KuhlPRB2010}, or ultracold atoms in hexagonal optical
lattices \cite{soltan-hexa, leticia-ag}. The latter system offers
a great deal of control, and several pioneering experiments have
proven the versatility of cold atoms as quantum simulators during
the past decade, cf. Ref. \cite{mlbook}. By engineering
hexagonal optical lattices for bosons \cite{soltan-hexa, Jo2012,
DucaScience2015,LiScience2016, weinberg2016} and fermions
\cite{leticia-ag, UehlingerEPJB2013, uehlinger,
jotzu14,FlaeschnerScience2016}, cold atom systems with
graphene-like band structure have been realized in recent years.
These experimental possibilities have also stimulated theoretical
interest in atoms on hexagonal lattices \cite{duan-ag,
sols,LeePRA2009, poletti2011,luehmann2014,cao2015}. A particular
feature of cold atoms in optical lattice is the tunability of the
tunneling strengths. In the context of artificial graphene, this
allows to smoothly introduce anisotropies, which can result into
the merging of the Dirac points, and mass acquisition of the
excitation \cite{duan-ag, sols,LeePRA2009, leticia-ag}. Furthermore, a gap at the Dirac points can be opened by introducing an energy offset between the two sublattices as studied experimentally in Refs. \cite{leticia-ag,weinberg2016,uehlinger}, or, more recently, by applying an artificial staggered magnetic field (realizing the celebrated Haldane model) \cite{jotzu14}.

 \begin{figure}[t]
\centering
\includegraphics[width=0.70\textwidth, angle=0]{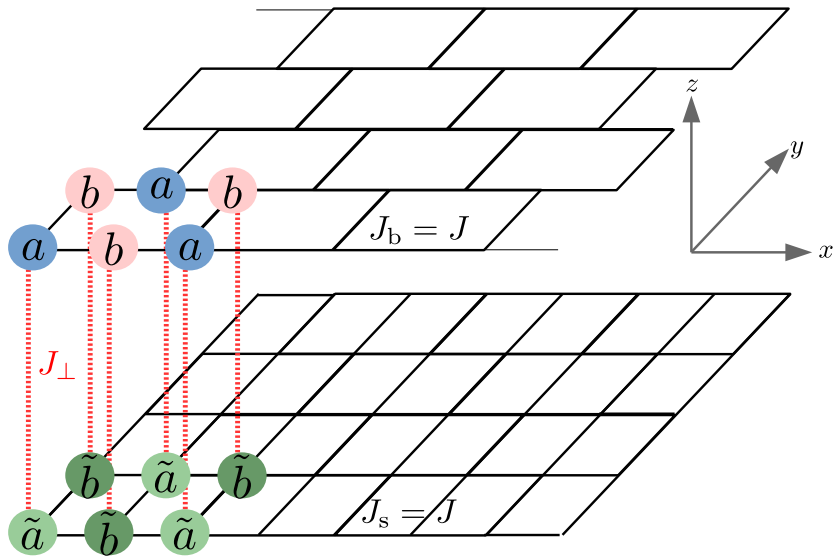}
\caption{\label{scheme} Schematic view of the
bilayer system: On top of a square lattice we have a
brick-wall layer, with a tunable coupling $J_\perp$ between the layers.
The brickwall lattice can be divided into two sublattices $a$ and $b$. The coupling carries this bipartite structure over to the square lattice, where we distinguish between operators $\tilde a$ and $\tilde b$.
}
\end{figure}

In the present paper we propose a different way for manipulating
the band structure of artificial graphene of cold atoms in optical
lattices. Motivated by the so-called ``proximity effect'' where
different materials are placed together in order to engineer
topological superconductors
\cite{lutchyn2010,sau2010,hui2015,cole2016}, we consider a
heterostructure of two layers with different dispersion. This
scenario is closely related to the approaches recently used in
real-graphene systems, where the proximity between graphene to
other substrates is exploited both theoretically and
experimentally. The proximity to hexagonal boron nitride has been
studied in Refs. \cite{ortix2012,yankowitz2012}. A graphene
layer close to a semi-conducting quantum well was considered in
Refs. \cite{principi12,gamucci14,aliaj16}, studying the
interplay of a two-dimensional electron gas with a layer of Dirac
fermions. Such scenario is relevant also in the context of a
recent experiment which implements an artificial-graphene layer on
top of a semiconductor \textit{via} electron-beam lithography
\cite{wind15,wang16}. A somewhat similar setup is considered in
the present paper. We theoretically study a bilayer structure,
where one layer has a graphene-like band structure, while the
other layer has square lattice geometry. As illustrated in Fig.
\ref{scheme}(a), the ``graphene'' layer is realized by a brick-wall
lattice, which regarding its band structure is equivalent to a
hexagonal lattice. With this, the square layer perfectly matches
the graphene lattice with respect to the atom positions. It
differs, however, with respect to the coupling between the atoms,
as in the brick-wall layer every second link along the
$y$-direction is missing. We then investigate the role of a
coupling between the layers, for a non-interacting system in Sec.
\ref{sec:band}, for a system with weak attractive interactions in
Sec. \ref{sec:super}, and for a system with strong repulsive
interactions in Sec. \ref{sec:mott}. We describe a scheme for
realizing such system with cold atoms in optical lattices in Sec.
\ref{sec:oli}.

The bilayer system here proposed manifests a rich set of physical
phenomena. Our main findings are summarized as follows:

(i) In the non-interacting system, the number and position of
Dirac points are controlled by the interlayer coupling. In the presence of interlayer coupling two additional Dirac points arise from intersections
between brick-wall and square lattice dispersion. For strong
coupling, the Dirac points stemming from the brick-wall dispersion merge and disappear. If prepared at half
filling, the system then undergoes a transition from a semi-metal
to a band insulator. At this transition, the lowest excitations are massless in one direction, but massive in the other direction.

(ii) For attractive interactions we find, within a mean-field
approximation, a competition between superfluid and semi-metallic phases. In contrast to a single brick-wall layer, in the bilayer such competition also occurs at filling 1/4 and 3/4, due to the presence of additional Dirac points.

(iii) In the strongly repulsive system, for which the low-energy physics is described by an appropriate spin model, we have used the quantum Monte Carlo method in order to investigate the competition between long-range order within each layer, and dimerization of nearest-neighbors between the two layers. We have quantitatively determined the critical interlayer coupling at which a valence bond crystal is formed ($J_\perp/J=2.2$).

\section{System}
As illustrated in Fig. \ref{scheme}, we consider a bilayer
structure consisting of one square lattice on top of a brick-wall
lattice. The brick-wall lattice mimics graphene, as it represents
a deformed hexagonal lattice. Accordingly, we can divide it into
two triangular sub-lattices, $A$ and $B$, and the annihilation
operators on these sub-lattices are denoted by $a_{\bf i}$
and $b_{\bf i}$. Without coupling between the layers the unit cell of a square lattice contains only one atom. However, the presence of a finite coupling increases the unit cell to two atoms per layer, and we need to distinguish between the two sublattices also in the square layer. Accordingly, we denote the operators which annihilate particles
in the square lattice by $\tilde a_{\bf i}$ and $\tilde b_{\bf i}$. The index $\bf i$
denotes the position of the annihilated particle within the
$xy$-plane. With these definitions, and assuming isotropic
couplings $J_{\rm s}$ and $J_{\rm b}$ in the square and the
brick-wall lattice, respectively, the tight-binding Hamiltonian of
the bilayer system reads
\begin{align}
H_{\rm tb} =& -J_{\rm b} \sum_{{\bf i }\in A} \left(
b_{{\bf i}+\hat x}^\dagger a_{\bf i} + b_{{\bf i}-\hat x}^\dagger a_{\bf i} + b_{{\bf i}+\hat y}^\dagger a_{\bf i} \right)
 - J_{\rm s} \sum_{\langle {\bf i},{\bf j}\rangle} \tilde b_{\bf j}^\dagger \tilde a_{\bf i}  
 - J_{\perp}  \sum_{{\bf i}} \left( a_{\bf i}^\dagger \tilde a_{\bf i} + b_{{\bf i}}^\dagger \tilde b_{{\bf i}}\right) + {\rm H.c.},
 \end{align}
 where $J_{\perp}$ is the interlayer coupling. The vectors $\hat x$ and $\hat y$ connect neighboring sites along the $x$- and $y$-direction.  The sum over $\langle \bf i,j \rangle $ contains all nearest-neighbor pairs with $\bf i$ in sublattice $A$ and $\bf j$ in sublattice $B$. 
The last sum, over $\bf i$, shall include all lattice points in the corresponding sublattice, i.e. it is restricted to sublattice $A$ for terms containing $a,\tilde a$ operators, and restricted to sublattice $B$ for terms containing $b,\tilde b$ operators.

 Loading the lattice with spin-polarized fermions, no double-occupancies can occur, and interactions are avoided. In this scenario, further investigated in the next section, we can explore the band structure of the bilayer. Interesting many-body effects, considered in the subsequent sections, will occur when the lattice is loaded with fermionic atoms having a (pseudo)spin-1/2 degree of freedom. In this case, we can have up to two atoms on each site, and local interactions become important. The Hamiltonian then reads $H=\sum_{\sigma=\uparrow,\downarrow} H_{{\rm tb},\sigma} + H_{\rm int}$, where the $\sigma$ index denotes the additional spin degree of freedom carried by each particle operator. The interaction Hamiltonian $H_{\rm int}$ is given by
\begin{align}
\label{HU}
 H_{\rm int}= U \sum_{\bf i} (a_{{\bf i}\uparrow}^\dagger a_{{\bf i}\downarrow}^\dagger a_{{\bf i}\downarrow} a_{{\bf i}\uparrow} + b_{{\bf i}\uparrow}^\dagger b_{{\bf i}\downarrow}^\dagger b_{{\bf i}\downarrow} b_{{\bf i}\uparrow} + \tilde a_{{\bf i}\uparrow}^\dagger \tilde a_{{\bf i}\downarrow}^\dagger \tilde a_{{\bf i}\downarrow} \tilde a_{{\bf i}\uparrow}  + \tilde b_{{\bf i}\uparrow}^\dagger \tilde b_{{\bf i}\downarrow}^\dagger \tilde b_{{\bf i}\downarrow} \tilde b_{{\bf i}\uparrow}),
\end{align}
where we assume equal interaction strengths $U$ in both layers. Again the sum over $\bf i$ is assumed to cover all points in the corresponding sublattice.

\section{Band structure of the lattice \label{sec:band}}

\begin{figure*}
\centering
\includegraphics[width=0.98\textwidth, angle=0]{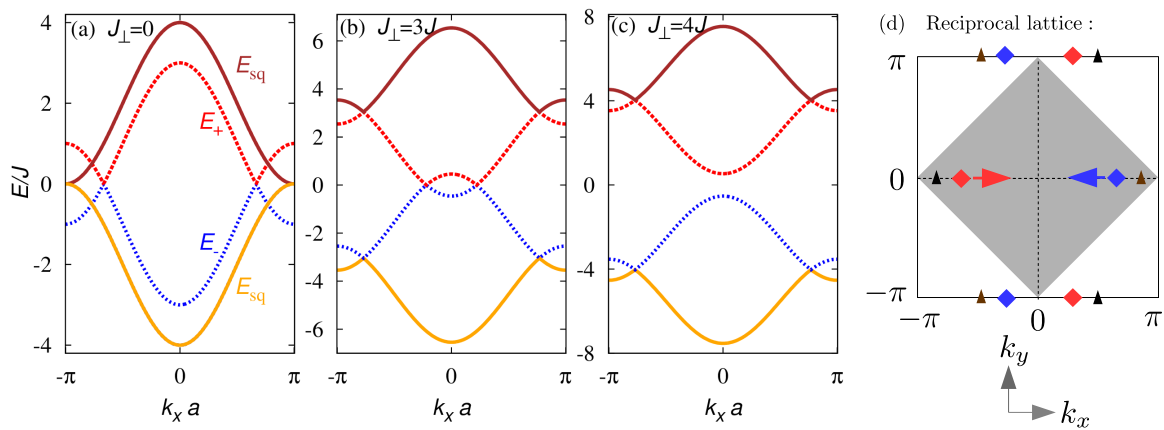}
\caption{\label{Fig2} Dispersion of the two-layer lattice along $k_x$ for different interlayer coupling strengths:
(a) no coupling between the layers ($J_\perp=0$), (b) $J_\perp=3J$, (c) $J_\perp=4J$.
All dispersions are obtained at $k_y=0$. In (a), we distinguish between the brick-wall lattice dispersions $E_\pm$ (red and blue) with two Dirac points at ${\bf k}=(\pm 2\pi/3,0)/a$, and the square-lattice dispersion $E_{\rm sq}$, split into two bands (brown and orange). At finite coupling, (b) and (c), Dirac points arise at the intersections between brick-wall and square layer, ${\bf k}=(\pm{\rm acos}(-3/4),0)/a$. The interlayer coupling moves the brick-wall Dirac points towards the center of the Brillouin zone. In (c), the brick-wall Dirac points have disappeared. In (d), we show a sketch of the reciprocal lattice, with the shaded region being the first Brillouin zone of the coupled bilayer. The rhombs mark the positions of the brick-wall Dirac points, with the arrows marking their direction upon increasing the coupling. The triangles denote the positions of the Dirac points due to intersecting bands.
}
\end{figure*}

Due to translational invariance in the $xy$-plane, the eigenstates
of $H_{\rm tb}$ are characterized by their wave vector ${\bf
k}=(k_x,k_y)$. The real-space unit cell is spanned by vectors $(\hat x,\pm  \hat y)$, and contains two sites per layer. This corresponds to a Brillouin zone as shown in Fig. \ref{Fig2}, identical to the Brillouin zone of a brick-wall lattice, but half of the size of the square-lattice Brillouin zone. Reciprocal lattice vectors connecting equivalent points in ${\bf k}$-space are spanned by $\Delta {\bf k}_\pm=(\pi,\pm \pi)/a$. Given two sublattices in each of the two layers, there are in total four eigenstates at each $\bf k$, that is, four bands form the band structure of the lattice. As we elaborate in this section, the band structure undergoes a topological transition upon tuning the interlayer coupling strength $J_{\perp}$.

We start by considering the uncoupled system, $J_{\perp} \rightarrow 0$. The band structure then consists of the two
graphene bands $E_\pm({\bf k})$, and the band of the square lattice $E_{\rm sq}({\bf k})$:
\begin{align}
E_\pm ({\bf k}) =& \pm E_{\rm br}({\bf k}) = \pm J \sqrt{ 3 + 2 \cos(2k_x a)  + 2\cos[(k_x+k_y)a]
 + 2\cos[(k_x - k_y)a] },
\\
E_{\rm sq}({\bf k}) =& - 2J \left[ \cos(k_x a) +\cos(k_y a) \right],
\end{align}
with $a$ the lattice spacing, set to 1 in the following. For
simplicity, we choose $J_{\rm s}=J_{\rm b}=J$. Naturally the square-layer band extends over a larger Brillouin than the brick-wall bands, but we can shift any point of the reciprocal lattice into the first Brillouin zone by $\Delta {\bf k}_\pm$. This shift splits the square-layer dispersion into two branches. A cross section of the band structure along $k_y=0$ is plotted in Fig.\ref{Fig2}(a). The bands from the brick-wall lattice $E_-$ and $E_+$, depicted in blue and red, respectively, touch each
other in Dirac points at momenta ${\bf K}_{\pm}=(\pm 2\pi/3,0)/a$. Within the first Brillouin zone, we have a single pair of Dirac points at zero energy. Of course, the bands also look
gapless at conical intersections between the square lattice band
(brown and orange solid line in Fig. \ref{Fig2}(a)) and the brick-wall
lattice bands. However, it is important to note that no gapless
excitations live at these intersections. A gapless excitation
would correspond to a particle hopping from the brick-wall to the
square lattice or vice versa, which is impossible for
$J_{\perp}=0$.

This is qualitatively different in the presence of a finite coupling between the layers: A gap opens at the conical intersections, except for two separated points in ${\bf k}$-space. These points now provide additional Dirac points, with gapless excitations, and linear dispersion along any direction in $\bf k$-space. We have also calculated the Berry phase $\gamma$ for each of the bands $\varphi_i({\bf k})$ along a contour encircling these intersection points, $\gamma=\langle \varphi_i({\bf k}) | \Delta_{\bf k}  \varphi_i({\bf k})\rangle$. At any finite $J_\perp$, we obtain $\gamma=\pi$, thus the intersection points indeed are topologically protected Dirac points. We also note that the position of the intersection points is independent from the coupling $J_\perp$, given by ${\bf k}_{\rm IS\pm} = (\pm{\rm acos}(-3/4),0)/a$. At each of these two points, we have one band-touching at negative energy $E_-/J=-A(J_\perp)/2$, between the first and the second band, and one at positive energy $E_+/J=A(J_\perp)/2$, between the third and the fourth band. Here we have introduced the short-hand notation 
\begin{align}
\label{Ash}
A(J_\perp)=\sqrt{1+4(J_\perp/J)^2}.
\end{align}
In order to have these Dirac points at the Fermi surface, we need to adjust the filling factor to $1/4$ for the Dirac points at negative energy, or $3/4$ for the Dirac points at positive energy. At these filling factors, the non-interacting system exhibits a semi-metallic phase for any finite value of $J_\perp$. The Fermi velocity around these Dirac points is anisotropic. Along the $k_x$-axis, we have $E(k_x)=E_\pm \pm \hbar v_x k_x$, with $v_x/v_0 = \sqrt{7}/2$, where $v_0=J a/\hbar$ is a unit of velocity. Along the $k_y$-axis, we have $E(k_y)=E_\pm \pm \hbar v_y k_y$, with $v_y/v_0 = \frac{J_\perp/J}{A(J_\perp)}$.  

In contrast to the intersection points, the Dirac excitations directly originating from the brick-wall
lattice do not remain at fixed $\bf k$-positions upon changing the interlayer coupling, but their energy is pinned to $E=0$. At any $J_\perp$, they are located between the second and the third bad, i.e. they correspond to filling 1/2. While their $k_y$ position remains pinned at $k_y=0$, their $k_x$-position is parametrized as
\begin{align}
 k_x(J) = \pm \frac{1}{a} {\rm acos}\left( -\frac{3}{4}+\frac{1}{4}A(J_\perp) \right),
\end{align}
that is, they move towards the center of the Brillouin zone when $J_\perp$ is increased, see Fig. \ref{Fig2}(b). At a critical coupling $J_{\perp,\rm crit}=2\sqrt{3} J$, the two Dirac points merge at $k_x=0$. For even larger values of the coupling, a gap opens and the Dirac points disappear, as shown in \ref{Fig2}(c). The fact that the Dirac points can only merge at a high symmetry point of the Brillouin zone (here the center) is a generic feature, occurring as well in single-layered deformed honeycomb lattices \cite{sols,LeePRA2009,montambaux09}.

The non-interacting system at half filling thus undergoes a semi-metal to band insulator transition at $J_{\perp,\rm crit}$. Near criticality the system exhibits some remarkable, highly anisotropic features: In the semi-metallic phase excitations are characterized by the Fermi velocities, which along the two lattice axes read:
\begin{align}
 v_x(J_\perp)/v_0 &= \frac{1}{2}\sqrt{7-A(J_\perp)^2+6A(J_\perp)}, \\
 v_y(J_\perp)/v_0 &= \frac{1}{2}\left(1+\frac{1}{A(J_\perp)}\right),
\end{align}
for $J_\perp \leq J_{\perp,\rm crit}$. In the insulating phase the dispersion relation are massive $E(k_i) \sim \Delta(J_\perp) + \frac{\hbar^2 k_i}{2m_i(J_\perp)}$, and the excitations are characterized by their effective mass $m_i(J_\perp)$, which again depends on the lattice direction. We obtain
\begin{align}
 m_x(J_\perp)/m_0 &= 1/2, \\
 m_y(J_\perp)/m_0 &= \frac{7A(J_\perp) |7-A(J_\perp)|}{2[16+5A(J_\perp)^2-35A(J_\perp)]},
\end{align}
for $J_\perp \geq J_{\perp,\rm crit}$. Here we have introduced $m_0=\hbar^2/(J a^2)$ as a unit of mass. The insulating gap $\Delta(J_\perp)$ reads
\begin{align}
 \Delta(J_\perp)=\frac{1}{2}\left|7-A(J_\perp)\right|.
\end{align}

\begin{figure*}
\centering
\includegraphics[width=0.98\textwidth, angle=0]{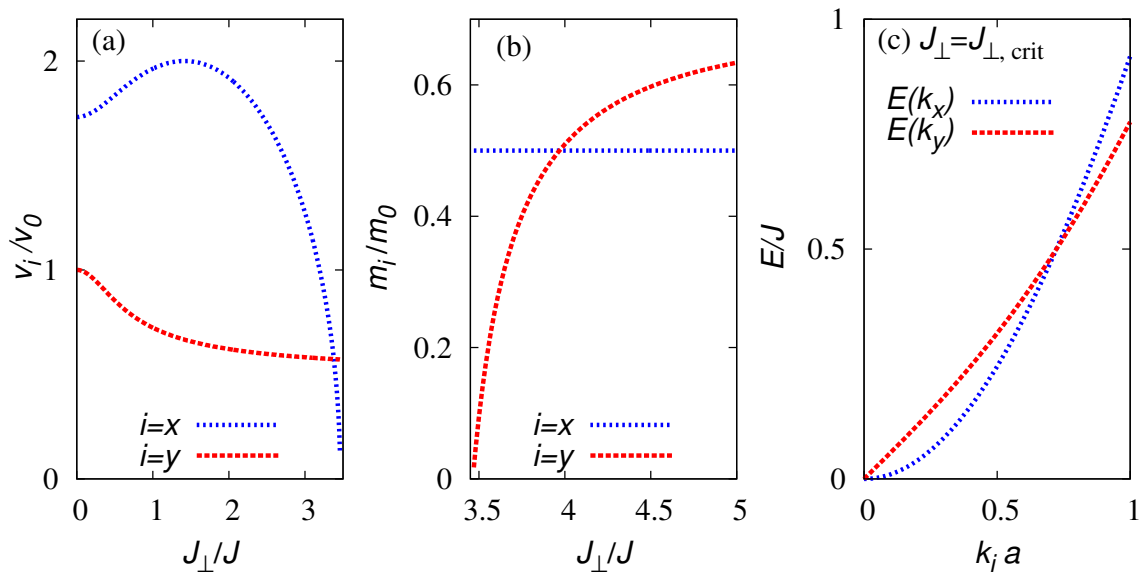}
\caption{\label{fig:crit} Behavior of excitation near the Dirac point merging: Fermi velocities (a) and effective masses (b) characterize the excitations above/below the critical coupling. They are anisotropic, and while the Fermi velocity along $x$ vanishes at the critical point, the excitations along $y$ remain massless at the critical point. The coexistence of massive and massless excitations at the critical point is illustrated by the energy spectrum in (c).
}
\end{figure*}

We have plotted the Fermi velocities and effective masses in Fig. \ref{fig:crit} (a) and (b). While $v_x$ decreases smoothly to zero when the critical coupling is reached, $v_y$ remains non-zero even at the critical point, and drops to zero abruptly for larger values of $J_\perp$. On the other hand, the effective mass along $y$ is zero at the critical point, and continuously increases for larger $J_\perp$, while $m_x$ takes a constant finite value for any $J_\perp \geq J_{\perp,\rm crit}$. At the critical point, this gives rise to the coexistence of massless excitations along $k_y$, and massive excitations along $k_x$, as illustrated in Fig. \ref{fig:crit} (c). Such behavior has also been predicted in single-layered graphene-like systems \cite{montambaux09,dietl08,leticia-ag}

Experimentally, the existence and position of Dirac points can be probed by
accelerating the atomic cloud \textit{via} a magnetic field gradient. If the atoms pass a
Dirac point, there is transfer between the bands, which becomes visible in the
quasi-momentum distribution after a Bragg reflection. This method has been
pioneered in Ref. \cite{leticia-ag}.

\section{Superfluidity in bilayer system with attractive interactions \label{sec:super}}

In the previous section, we considered the single particle (free) physics characterizing the coupled-layer system. This can be obtained by loading the lattice with spinless or fully spin-polarized fermionic atoms. In the following sections, we will consider the bilayer lattice loaded with two species of fermions, defining a pseudospin-1/2 degree of freedom (denoted by $\uparrow$ and $\downarrow$). Then, the fermions may interact, as described by the Hamiltonian $H_{\rm int}$. In the present section, we consider the case of weak \textit{ attractive} interactions ($U\equiv-u<0$).

It is known that for uncoupled layers a pairing instability,
captured by BCS theory, occurs in both the square and the
brick-wall lattice, giving rise to a superfluid phase. However,
while fermions in the square lattice form a superfluid at any
attractive interaction \cite{mlbook}, a peculiarity happens in the
brick-wall lattice. At half filling ($n=1$), the Dirac points are
located on the Fermi surface, and below a critical interaction
strength $u < u_{\rm crit}$, the system remains in a semi-metal
phase \cite{paramekanti06,paramekanti12}.

To make predictions for the coupled layers, we follow the
procedure of standard BCS theory, that is, we treat the
interacting Hamiltonian $H=\sum_\sigma H_{{\rm tb},\sigma}+H_{\rm
int}$ on a mean-field level. To connect our results to the
uncoupled case, we separately study the gap parameter on the
brick-wall lattice, $\Delta_{\rm br} \equiv (4u/N) \sum_{\bf k}
\langle a_{-{\bf k}\downarrow} a_{{\bf k}\uparrow} \rangle =
(4u/N) \sum_{\bf k} \langle b_{-{\bf k}\downarrow} b_{{\bf
k}\uparrow} \rangle$, and the gap parameter on the square lattice
$\Delta_{\rm sq} \equiv (4u/N) \sum_{\bf k} \langle \tilde a_{-{\bf
k}\downarrow} \tilde a_{{\bf k}\uparrow} \rangle = (4u/N) \sum_{\bf k} \langle \tilde b_{-{\bf
k}\downarrow} \tilde b_{{\bf k}\uparrow} \rangle$. In these expressions,
the factor $N/4$ stems from the number of sites per sublattice (with $N$ the total number of sites). The BCS
Hamiltonian then reads
\begin{align}
 H_{\rm BCS} & = \sum_{\sigma \neq \sigma'} \sum_{\bf k} \left( a_{{\bf k}\sigma}^\dagger, a_{-{\bf k}\sigma'}, b_{{\bf k}\sigma}^\dagger, b_{-{\bf k}\sigma'}, \tilde a_{{\bf k}\sigma}^\dagger, \tilde a_{-{\bf k}\sigma'}, \tilde b_{{\bf k}\sigma}^\dagger, \tilde b_{-{\bf k}\sigma'}  \right)
  \cdot \nonumber \\ &
  \left(
 \begin{matrix}
  -\mu & \Delta_{\rm br} & -J_{\bf k}^{\rm br} & 0 & -J_{\perp} & 0 & 0 & 0 \\
  \Delta_{\rm br}^* & \mu & 0 & J_{{\bf k}}^{\rm br} & 0 & J_{\perp} & 0 & 0\\
  -J_{-\bf k}^{\rm br} & 0 & -\mu & \Delta_{\rm br} & -0 & 0& -J_{\perp} & 0 \\
  0 & J_{-{\bf k}}^{\rm br} & \Delta_{\rm br}^* & \mu & 0 & 0 & 0 & J_{\perp} \\
  -J_{\perp} & 0 & 0 & 0 & -\mu & \Delta_{\rm sq} & -J_{\bf k}^{\rm sq} & 0 \\
  0 & J_{\perp} & 0 & 0 & \Delta_{\rm sq}^* & \mu & 0 & J_{\bf k}^{\rm sq} \\
  0 & 0 & -J_{\perp} & 0 & -J_{\bf k}^{\rm sq} & 0 & -\mu & \Delta_{\rm sq} \\
  0 & 0 & 0 & J_{\perp} & 0 & J_{\bf k}^{\rm sq} & \Delta_{\rm sq}^* & \mu \\
 \end{matrix}
 \right)
 \cdot
 \left(
 \begin{matrix}
 a_{{\bf k}\sigma} \\
 a_{-{\bf k}\sigma'}^\dagger \\
 b_{{\bf k}\sigma} \\
 b_{-{\bf k}\sigma'}^\dagger \\
 \tilde a_{{\bf k}\sigma} \\
 \tilde a_{-{\bf k}\sigma'}^\dagger\\
 \tilde b_{{\bf k}\sigma} \\
 \tilde b_{-{\bf k}\sigma'}^\dagger
\end{matrix}
\right)
\nonumber \\ & \equiv \sum_{{\bf k}\sigma} \alpha_{{\bf k}\sigma}^\dagger M_{{\bf k}\sigma} \alpha_{{\bf k}\sigma}.
\end{align}
In this notation we have introduced $J_{\bf k}^{\rm br} \equiv 2J
\cos(k_x) + J {\rm e}^{-i k_y}$, and $J_{\bf k}^{\rm sq}=2J
\left[\cos(k_x)+\cos(k_y)\right]$, with $J$ the intralayer hopping
strength. Note that $J_{-{\bf k}}^{\rm br}=(J_{\bf k}^{\rm
br})^*$, such that the matrix is Hermitian. The parameter $\mu$ is
a shifted chemical potential, which shall absorb the corresponding
mean-field terms $U\langle a_{{\bf k}s}^\dagger a_{{\bf k}s}
\rangle$, $U\langle b_{{\bf k}s}^\dagger b_{{\bf k}s} \rangle$,
$U\langle \tilde a_{{\bf k}s}^\dagger \tilde a_{{\bf k}s} \rangle$, and $U\langle \tilde b_{{\bf k}s}^\dagger \tilde b_{{\bf k}s} \rangle$.
We simplify our analysis by asking that $\mu$ takes identical values in both
layers. With this restriction, we can only control the total
particle number, but not the number of particles in each layer. We
also disregard a possible spin dependence of $\mu$, which renders
the Hamiltonian symmetric under spin flips. We can therefore
neglect the summation over $\sigma$.

The Hamiltonian $H_{\rm BCS}$ is particle-hole symmetric, ${\cal P}H_{\rm BCS}{\cal P} = - H_{\rm BCS}$, where ${\cal P}$ denotes particle-hole conjugation. Thus, the energy spectrum is symmetric around $E=0$. The eigenstates at positive energy describe quasi-particle excitations, and the many-body ground state is the quasi-particle vacuum $\ket{\Phi}$. We denote the quasi-particle operators by $\nu_{i,{\bf k}\sigma}$, where the different types of excitations are captured by the index $i$ running from 1 to 3. The corresponding energies are denoted $E_{i,{\bf k}}$. In this notation the Hamiltonian reads:
\begin{align}
 H_{\rm BCS} = \sum_{\sigma,{\bf k}} \sum_{i=1}^3 E_{i,\rm k} \nu_{i,{\bf k}\sigma}^\dagger \nu_{i,{\bf k}\sigma}.
\end{align}
Explicitly, the quasi-particle operators are given by
\begin{align}
 \nu_{i,{\bf k}\uparrow} = &  u_{1,{\bf k}\uparrow}^{(a)} a_{\bf k \uparrow} + v_{1,{\bf -k}\downarrow}^{(a)} a_{\bf -k \downarrow} + u_{1,{\bf k}\uparrow}^{(b)} b_{\bf k \uparrow}
 + v_{1,{\bf -k}\downarrow}^{(b)} b_{\bf -k \downarrow} + u_{1,{\bf k}\uparrow}^{(\tilde a)} \tilde a_{\bf k \uparrow} + v_{1,{\bf -k}\downarrow}^{(\tilde a)} \tilde a_{\bf -k \downarrow} \nonumber \\ &
  + u_{1,{\bf k}\uparrow}^{(\tilde b)} \tilde b_{\bf k \uparrow} + v_{1,{\bf -k}\downarrow}^{(\tilde b)} \tilde b_{\bf -k \downarrow},
\end{align}
and similarly for $\nu_{i,{\bf k}\downarrow}$, with the coefficients $u_{i, {\bf k}\sigma}$ and $v_{i, {\bf k}\sigma}$ obtained from diagonalization of $M_{{\bf k}\sigma}$.

Once these coefficients are known, we can evaluate the number of
particles and the pairing gap. By rewriting the original operators
in terms of quasi-particle operators, it is straightforward to
evaluate expectation values with respect to quasi-particle states.
In the simplest case, at zero temperature, we just have to take
the average $\langle \cdot \rangle$ with respect to the
quasi-particle vacuum $\ket{\Phi}$. We obtain:
\begin{align}
\label{aa}
  \langle a_{{\bf k}\uparrow} a_{-{\bf k}\downarrow} \rangle = u_{1,{\bf k}\uparrow}^{(a)} v_{1,-{\bf k}\downarrow}^{(a)*} + u_{2,{\bf k}\uparrow}^{(a)} v_{2,-{\bf k}\downarrow}^{(a)*} + u_{3,{\bf k}\uparrow}^{(a)} v_{3,-{\bf k}\downarrow}^{(a)*} + u_{4,{\bf k}\uparrow}^{(a)} v_{4,-{\bf k}\downarrow}^{(a)*},
\end{align}
and the same expressions for $\langle b_{{\bf k}\uparrow} b_{-{\bf k}\downarrow} \rangle$, $\langle \tilde a_{{\bf k}\uparrow} \tilde a_{-{\bf k}\downarrow} \rangle$, and $\langle \tilde b_{{\bf k}\uparrow} \tilde b_{-{\bf k}\downarrow} \rangle$ with the upper indices on the right-hand-side replaced accordingly.
Similarly, we have
\begin{align}
\label{na}
 n_{\bf k}^{(a)} & \equiv  \langle a_{{\bf k}\uparrow}^\dagger a_{{\bf k}\uparrow} + a_{-{\bf k}\downarrow}^\dagger a_{-{\bf k}\downarrow} \rangle
 = 2 \left( \left|v_{1,-{\bf k}\downarrow}^{(a)} \right|^2
 + \left|v_{2,-{\bf k}\downarrow}^{(a)} \right|^2 + \left|v_{3,-{\bf k}\downarrow}^{(a)} \right|^2 + \left|v_{4,-{\bf k}\downarrow}^{(a)} \right|^2 \right),
\end{align}
and the corresponding expressions for $b$, $\tilde a$, and $\tilde b$.

The equations (\ref{aa}) and (\ref{na}), and their analogs for $b$, $\tilde a$, and $\tilde b$ can then be inserted into the gap equations. Restricting ourselves to zero temperature, we have
\begin{align}
\label{Dbr}
 \Delta_{\rm br} &= \frac{4u}{N}  \sum_{{\bf k} \in {\rm BZ}}
 \langle a_{{\bf k}\uparrow} a_{-{\bf k}\downarrow} \rangle = \sum_{{\bf k} \in {\rm BZ}} \frac{4u}{N} \langle b_{{\bf k}\uparrow} b_{-{\bf k}\downarrow} \rangle,
\\
\label{Dsq}
  \Delta_{\rm sq} &= \sum_{{\bf k} \in {\rm BZ}} \frac{4u}{N} \langle \tilde a_{{\bf k}\uparrow} \tilde a_{-{\bf k}\downarrow} \rangle = \sum_{{\bf k} \in {\rm BZ}} \frac{4u}{N} \langle \tilde b_{{\bf k}\uparrow} \tilde b_{-{\bf k}\downarrow} \rangle,
\end{align}
and the number equation for the filling $n$ per site
\begin{align}
 n = \frac{1}{N} \sum_{{\bf k} \in {\rm BZ}} \left( n_{\bf k}^{(a)} + n_{\bf k}^{(b)}+ n_{\bf k}^{(\tilde a)} + n_{\bf k}^{(\tilde b)} \right).
\end{align}
Here, the summations $\sum_{{\bf k}\in{\rm BZ}}$ shall cover the first Brillouin zone of the bilayer [see \ref{Fig2} (d)].

 \begin{figure}[t]
\centering
\includegraphics[width=0.8\textwidth, angle=0]{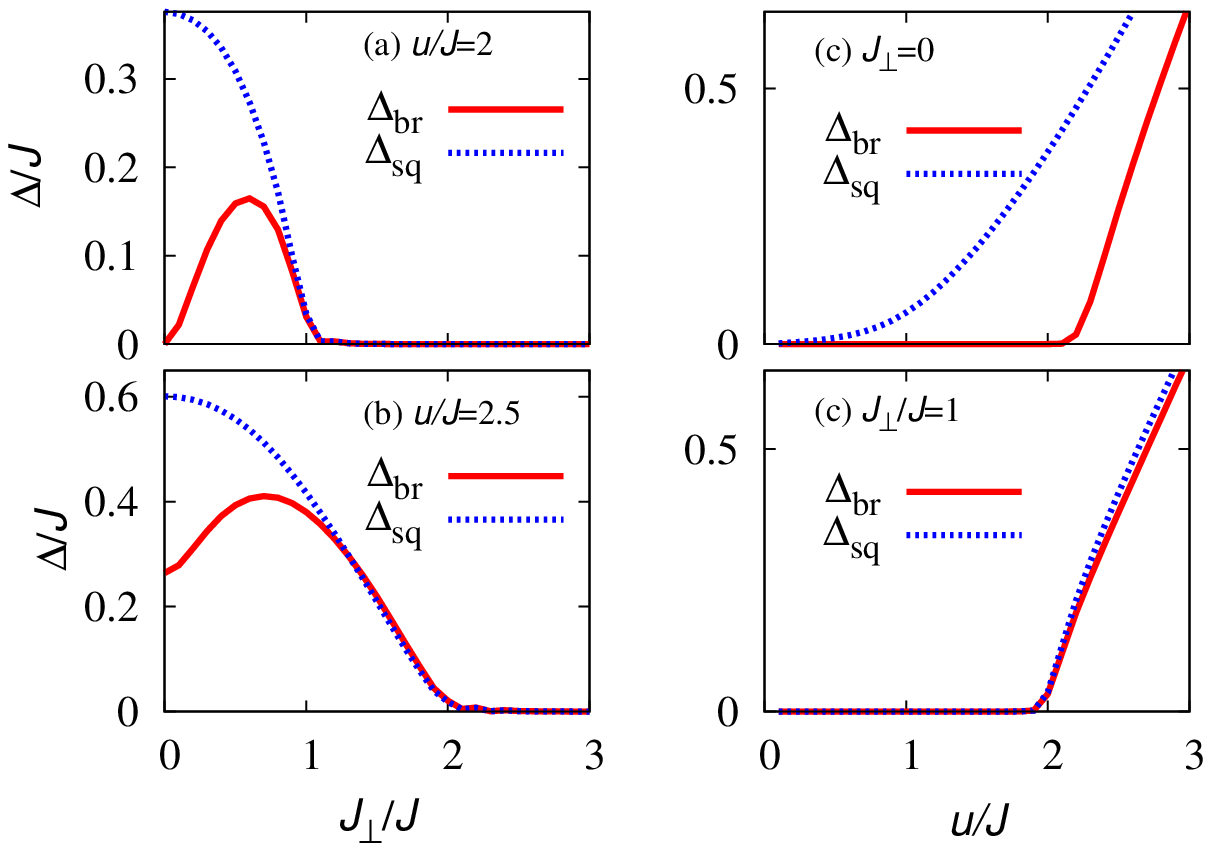}
\caption{\label{fig:gap_n1} Pairing gaps $\Delta_{\rm br}$ and
$\Delta_{\rm sq}$ in the brick-wall and the square layer at half
filling $n=1$. In (a,b), the gaps are plotted as a function of
$J_\perp$, while interactions $u=-U>0$ are kept
constant. In (c,d), we plot the gaps as a function of $u$ at fixed
values of $J_\perp$. }
\end{figure}

 \begin{figure}
\centering
\includegraphics[width=0.8\textwidth, angle=0]{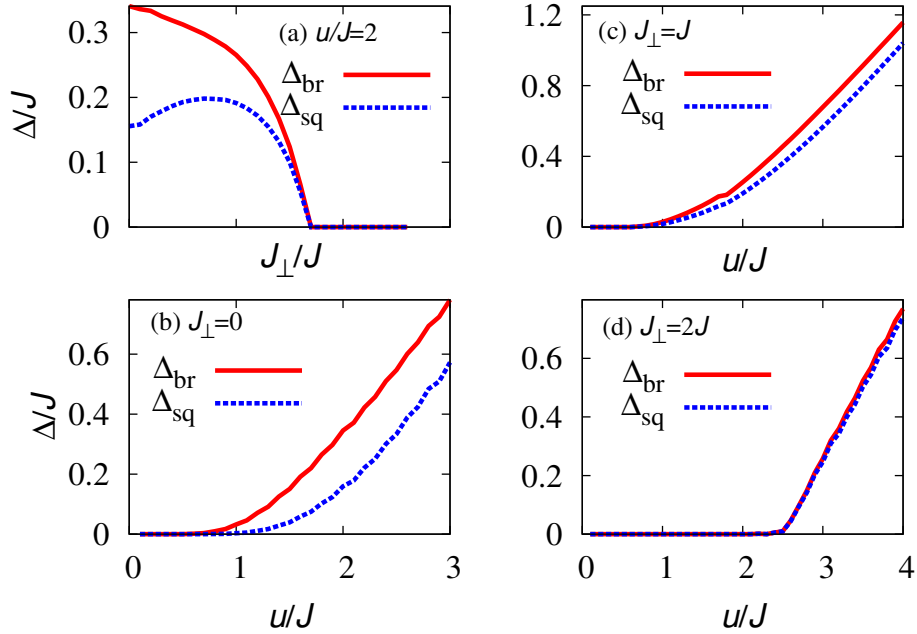}
\caption{\label{fig:gap_025} Pairing gaps $\Delta_{\rm br}$ and
$\Delta_{\rm sq}$ in the brick-wall and the square layer at
filling one-fourth, $n=1/2$. In (a) the gaps are plotted as a
function of $J_\perp$, while interactions $u$ are kept constant.
In (b--d), we plot the gaps as a function of $u$ at fixed values of
$J_\perp$. }
\end{figure}

We first consider a system at half filling ($n=1$), for which the
number equation is solved by setting $\mu=0$,  independently of
the gap parameters $\Delta_{\rm br}$ and $\Delta_{\rm sq}$. The
gap equations are then solved iteratively. The results are
illustrated in Fig. \ref{fig:gap_n1}. Let us first comment on
panel (c) which shows the behavior for two uncoupled layers,
$J_\perp=0$. As expected, the square lattice exhibits a finite
gap, $\Delta_{\rm sq}>0$, at any finite value of $u$. In contrast,
the gap in the brick-wall layer remains zero up to a critical
interaction strength $u_{\rm crit}\approx 2.1$ (in units of $J$
used throughout the discussion). This behavior corresponds to the
previously mentioned quantum phase transition from a semi-metal to
a superfluid phase. Panel (d) of Fig. \ref{fig:gap_n1} shows the
gaps as a function of $u$ for a finite value of $J_\perp$. Here, both layers behave approximately the same, exhibiting a semi-metal to superfluid transition. In comparison to the transition in a single brick-wall layer, the critical interaction strength is slightly reduced, $u_{\rm crit}\approx 1.9$. So at the same time, the interlayer coupling suppresses superfluid correlations in the square layer, and favors superfluid correlations in the brick-wall layer.

This interpretation can also be drawn from panels (a) and (b) of
Fig. \ref{fig:gap_n1}, showing the gaps for fixed values of $u$ as
a function of $J_\perp$. In (a), with $u=2$, the brick-wall layer
is in the semi-metal phase in the limit of $J_\perp=0$, but a
finite coupling immediately establishes a finite pairing gap in
both layers. Up to a certain value of $J_\perp$, the brick-wall
layer gap is further enhanced by the coupling, while the square
layer gap always decreases with $J_\perp$. For large $J_\perp$,
both gaps become strongly suppressed. The same behavior is seen also in panel (b) at
$u=2.5$, with only one qualitative difference seen in the
uncoupled case, $J_\perp =0$. At this interaction strength, also
the uncoupled brick-wall layer is superfluid.

In summary, the half-filled bilayer behaves qualitatively the same way as a half-filled brick-wall lattice. In both cases, the Dirac point at the Fermi surface causes a semimetal to superfluid transition at finite $U$. One might ask which role is played by the additional Dirac points at non-zero energy. Therefore, we first notice that they are at the Fermi surface for filling $n=1/2$ or filling $n=3/2$. Here we shall note that, due to particle-hole symmetry, the fillings $n<2$ and $2-n$ are equivalent. We also
note that, away from $n=1$, solving gap and number equations self-consistently is considerably more difficult, as also the chemical potential depends on the gap. 

Our main results are shown in Fig. \ref{fig:gap_025}. In the absence of interlayer coupling, shown in panel (b), both layers exhibit superfluid correlations for any non-zero strength of attractive interactions. Panel (a) shows that, upon turning on a sufficiently strong interlayer coupling, these superfluid correlations are suppressed. Interestingly, while for intermediate couplings, superfluid correlations can still be observed for arbitrarily small $u$, cf. panel (c) with $J_\perp=1$, stronger couplings require a finite interaction strength $u$ in order to establish a superfluid gap. Such scenario is illustrated in panel (d) with $J_\perp=2$. Thus, for strong enough interlayer coupling, the system  again exhibits a semi-metal to superfluid transition at filling one-fourth, $n=1/2$.

To conclude this section, we should mention that at half filling, the superfluid correlations are exactly degenerate with charge density wave correlations in the attractive Hubbard model on bipartite lattices (such as e.g. the lattice in this paper).
This is most easily seen by performing a particle-hole transformation on one species of spins, say $c_{i\downarrow} \mapsto \exp(iQ\cdot r_i) c_{i\downarrow}^\dagger$, where $Q=(\pi,\pi)$).  This transformation converts the attractive Hubbard model at arbitrary filling into the repulsive Hubbard model at half filling in a magnetic field that couples to the $z$-component of the spin \cite{PhysRevB.23.1447,PhysRevB.24.1579,PhysRevB.24.4018,PhysRevLett.66.946,ho09}. The superfluid correlations in the attractive model then map onto spin correlations in the $xy$ plane of the repulsive model, while charge density wave correlations map on to spin correlations in the $z$-direction. Now, if also the attractive Hubbard model is at half filling, the effective magnetic field in the repulsive model is zero, revealing an additional SU(2) symmetry of the original model. This rotational symmetry implies the degeneracy between superfluid and density wave phase. Hence, at half filling, even though we have only considered superfluid correlations, one must keep in mind that the ground state has the peculiar property of simultaneous long range phase coherence (superfluid) and density order.  

The charge density wave - superfluid state degeneracy is however broken on moving away from half filling, that is for a non-zero effective magnetic field in the repulsive Hubbard model, with the superfluid state always appearing to be energetically favorable (with the density wave correlations becoming short ranged) on the square \cite{PhysRevLett.66.946,PhysRevB.47.7995} and honeycomb lattices \cite{PhysRevB.80.245118,paramekanti06}.  On the other hand, longer range, i. e. off-site, interactions may stabilize charge density wave order or induce phase separation \cite{PhysRevB.23.1447,PhysRevB.24.1579,PhysRevB.24.4018}. However we do not consider such interactions in the paper.  In light of the above, we have focused on the generic instabilities of the studied model, that is pairing.

\section{ The brick-wall/square bilayer Heisenberg antiferromagnet \label{sec:mott}}

In this section, we will study the bilayer system with repulsive interactions, $U>0$. At half filling, {\it i.e.} when the total number of particles $N_{\rm tot}$ is equal to the number of lattice sites $N$, the system is in a Mott insulating phase for sufficiently large $U$. In the perturbative limit, where $U \gg J,J_\perp$, the low energy physics of the system is described by an effective spin-1/2 effective isotropic antiferromagnetic Heisenberg Hamiltonian
\begin{gather}
 H =  \sum_{\langle {\bf i},{\bf j} \rangle}  J_{\bf ij}^{\rm ex} {\bf S}_{\bf i} \cdot
{\bf S}_{\bf j}.
\label{HAFM}
\end{gather}
The indices ${\bf i}$ now collect both the position of the spins in the $xy$ plane, and with respect to the layer.
The antiferromagnetic (superexchange) coupling $J_{\bf ij}^{\rm ex}$ arises from  the virtual hopping of fermions  in the sector of no doubly occupied states constituting the degenerate manifold of states describing the ideal Mott state corresponding to $U=\infty$. Explicitly, it reads
\begin{align}
J_{\bf ij}^{\rm ex} =
 \begin{cases}
 4J^2/U \equiv J^{\rm ex}& \text{between $a,b$ sublattices and in the square layer, } \\
 4J_\perp^2/U \equiv J^{\rm ex}_\perp& \text{between the two layers, } \\
    0 & \text{otherwise.}
\end{cases}
\end{align}

We recall that at sufficiently low temperature, the isotropic
Heisenberg antiferromagnet on hypercubic lattices is characterized
by long-range antiferromagnetic order
\cite{review-heisenberg-square}. Despite the very low temperature
scale set by the superexchange interactions, antiferromagnetic
correlations have recently been observed in several experiments
using Fermi gases in optical lattices
\cite{leticia2,hulet14,Greif2015,Parsons2016,Boll2016,Cheuk2016,Drewes2016}.
Bipartite lattices in dimensions $d=2$, such as square and
brick-wall lattices, exhibit ground state N\'{e}el order
\cite{Sandvik0,Castro}. Strictly 2-dimensional and quasi
2-dimensional layered Heisenberg antiferromagnets have been
routinely studied motivated in part due to  possible relevance to
the physics of high-Tc superconductivity. The square-lattice
bilayer Heisenberg antiferromagnet is interesting because it can
be driven through an order-disorder quantum phase transition
\cite{Sandvik-Scalapino,Ganesh} by increasing the inter-plane
coupling realizing a lattice model instance of the continuum field
theory physics described by the (2+1)-dimensional non-linear
$\sigma$ model. It was suggested that certain magnetic properties
of cuprate superconductors are reminiscent of those of a magnetic
state close to such a quantum critical point
\cite{HighTcAFM,Chubukov,Sokol}. On the other hand, such  spin
Hamiltonians are also relevant to the description of so-called
spin dimer compounds which are  magnetically disordered (pairs of
spins bind together into singlets)  but fascinatingly may be
driven to order in a magnetic field \textit{via} Bose-Einstein condensation
of magnons \cite{bec, GiamarchiNaturePhys2008}.

The inequivalent bilayer structure considered in this paper may be realized in an optical lattice setup as discussed in the next section, which warrants the study of the associated Heisenberg model \eqref{HAFM}.
For interplane coupling
$J_\perp^{\rm ex} = 0$, the system consists of two decoupled
N\'{e}el ordered planes, although here each plane is characterized
by a different value of the staggered magnetization. In the
thermodynamic limit, the staggered magnetization per site in the
square lattice is $m_{\rm sq} \approx 0.3070 $ (see Ref.
\cite{Sandvik0}) while the lower coordination number in the
brick-wall lattice leads to a moderate decrease to the value
$m_{\rm hc}\approx  0.2677$ (see Ref. \cite{Castro}) due to
enhanced effects of quantum fluctuations. Coupling the two
lattices leads to an antiferromagnetically ordered two-layer
system.  In the opposite limit of $J_\perp^{\rm ex} = \infty$,
neighboring spins across the two layers are coupled to form
isolated singlets or dimers. The excitation spectrum in the dimer
phase is gapped, while it is gapless in the ordered phase. We
restrict here to the evaluation of the critical coupling
separating these two phases of the system. We mention that the two
coupled square layers and two coupled brick-wall layers have been
studied before yielding values $J_\perp^{\rm ex} \approx
2.522J^{\rm ex}$ (see Refs. \cite{Sandvik-Scalapino,Wang}) and
$J_\perp^{\rm ex} \approx 1.645J^{\rm ex}$ (see Ref.
\cite{Ganesh}), respectively.

Since the system is bipartite, quantum Monte Carlo simulations can
be performed without a sign problem. We use the projector Monte Carlo sampling method
in the valence bond basis (we consider an even number of
sites per layer) introduced by Sandvik \cite{Sandvik-vbs}
to obtain the ground state characteristics. We consider two
coupled layers  (a square and a brick-wall layer)   containing
$L\times L$ sites with a square aspect ratio, so that the total
number of sites is $N=2L^2$. In order to ensure convergence of
observables to the ground state values, the projection length in
the algorithm is chosen to be as high as $2L^3$. Performing the
simulations in the valence-bond basis allows to access
rotationally invariant correlation functions of the system (see
\textit{e.g.} Ref. \cite{Lin}). In order to distinguish the
N\'{e}el from dimer ordered phase, we consider the N\'{e}el
structure factor (or square of staggered magnetization) of both
the full lattice as well as for each layer separately. The
staggered magnetization is defined as
\begin{gather}
 m_s = \frac{1}{N_{\rm S}} \sum_i \phi_i \mathbf{S_i},
 \label{mag}
 \end{gather}
where the sum is over either all lattice sites or those pertaining
only to a single layer and $N_{\rm S}$ is the number of spins in
the considered system ($N_{\rm S}=N$ for the whole system and
$N_{\rm S}=N/2$ when considering a single layer. The staggering
variable $\phi_i$ is a sublattice parameter -- on dividing the
total bipartite lattice in sublattices $a$ and $b$,  $\phi_i=+1$
for sites on one sublattice $\phi_i=-1$ for the other lattice. The
structure factor is then defined as
\begin{gather}
S(\pi,\pi) = \langle m_s^2 \rangle
\label{structure}
\end{gather}
corresponding to ordering at the the wave vector $\textbf{k}=(\pi,\pi)$.

In order to locate the quantum critical point, we perform finite-size scaling of $S(\pi,\pi)$. Spin-wave theory predicts \cite{Huse} the following finite size scaling of the magnetization for an ordered state:
\begin{gather}
 \langle m_s(L)^2 \rangle = \langle m_s(L \rightarrow\infty)^2 \rangle + \frac{A}{L} + \frac{B}{L^2} + \frac{C}{L^3} \ldots
\label{scaling}
\end{gather}
where $L$ is the linear size of the system. Similarly, chiral
perturbation theory and renormalization group calculations for the
non-linear $\sigma$ model yield the same form of finite size
scaling \cite{Sandvik0}. Hence this form can be used to estimate
the thermodynamic limit value of the staggered magnetization.

 \begin{figure}[t]
\centering
\includegraphics[width=0.48\textwidth, angle=0]{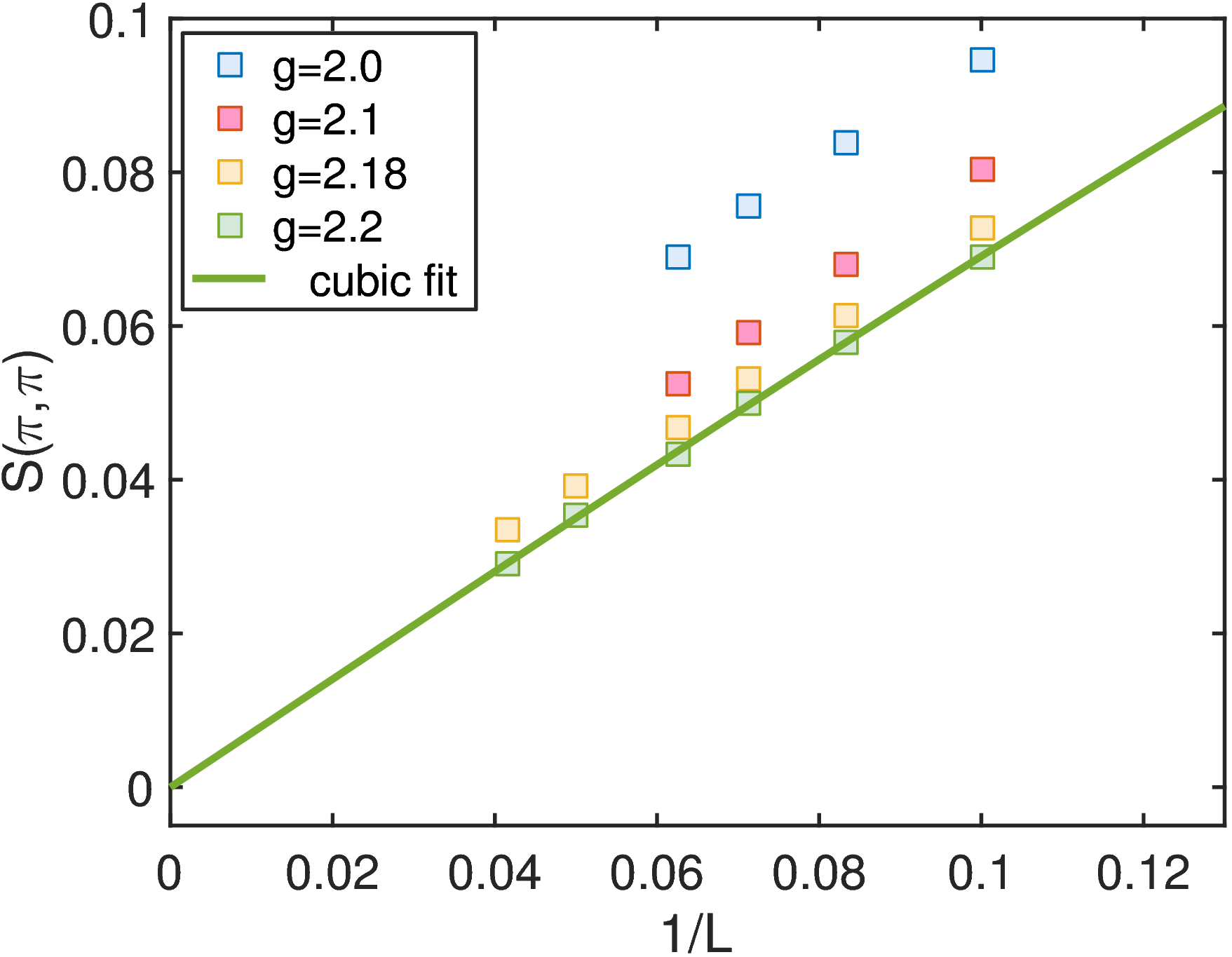}
\caption{\label{Fig3}
Scaling of the structure factor $S(\pi,\pi)$ of the bilayer lattice with inverse linear size of the system for various values of the control parameter $g=J_\perp^{\rm ex}/J^{\rm ex}$. The critical point is found near $g=2.2$, where $S(\pi,\pi)$ vanishes in the thermodynamic limit. Shown are data points for system sizes $L=10,12,14,16$ for all values of the control parameter, while close to the critical point $(g=2.18,2.20)$ data for $L=20,24$ are also plotted. The line is  a least square fit to the data with a polynomial of cubic degree in $1/L$.
}
\end{figure}

The scaling form \eqref{scaling} is well reflected in the obtained
Monte Carlo data for the two-layer magnetization data as seen in
Fig. \ref{Fig3}. Deep in the N\'{e}el phase, {\it e.g.} for the
inter- to intra-plane coupling ratio $g\equiv J_\perp^{\rm
ex}/J^{\rm ex}=2.0, 2.1 $ the dominant leading order finite size
correction is clearly of the order of $1/L$ for the system sizes
considered. On increasing $g$ the  higher-order corrections become
more important for lower system sizes. For $g=2.2$, we find a
polynomial of degree $3$ in the inverse systems size $1/L$ fits
the data well. The  value of the structure factor extrapolated to
the thermodynamic limit $L \rightarrow \infty$ is close to zero
for $g=2.2$. Hence we estimate the  quantum critical point to be
$g_c \approx 2.2$ for this model.

\begin{figure}[t]
\centering
\includegraphics[width=0.48\textwidth, angle=0]{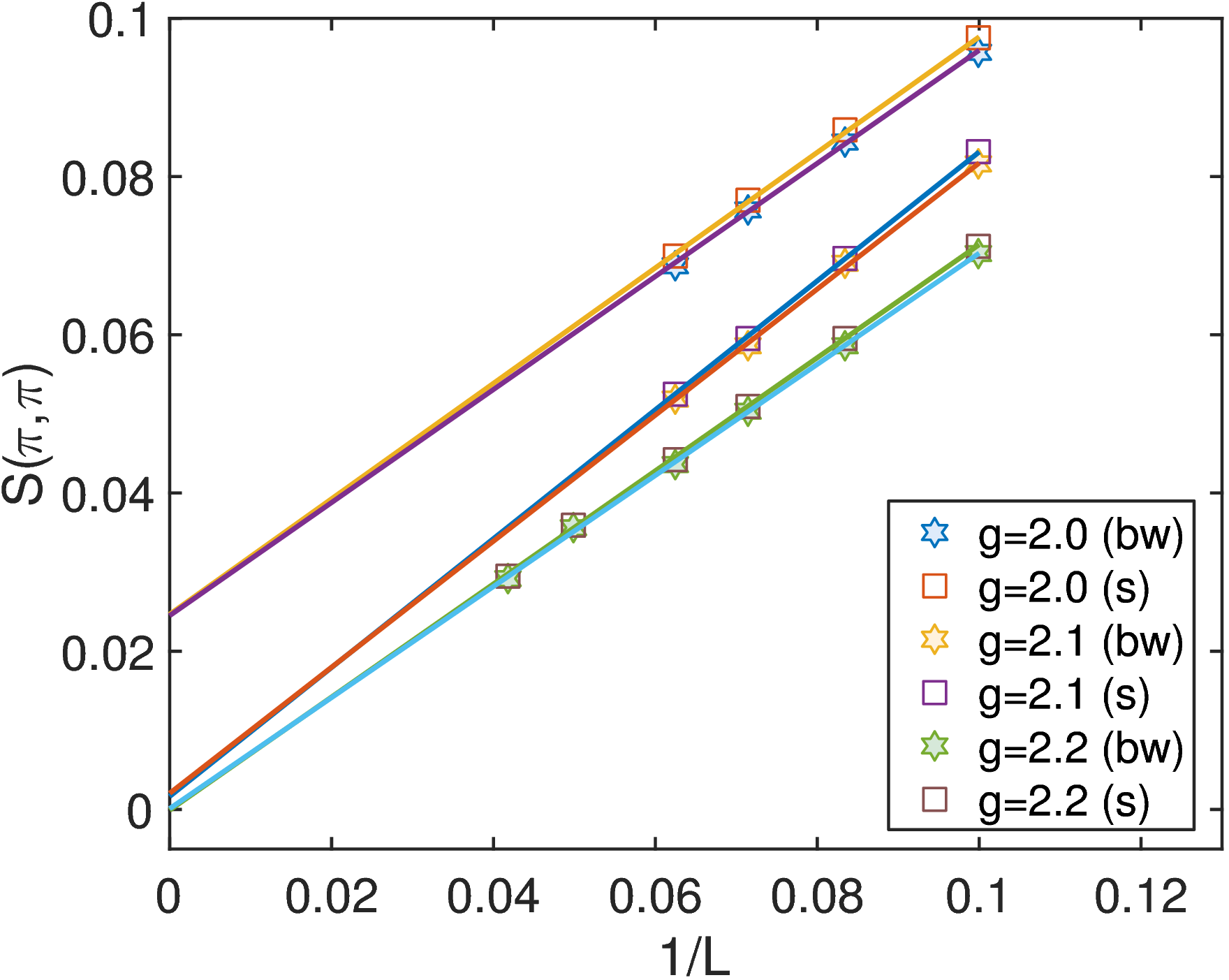}
\caption{\label{Fig4} In-plane structure factors. Shown are data
points for the same system sizes as in Fig. \ref{Fig3}, and
$g=2.0$, $2.1$, $2.2$ respectively. For each coupling strength
$g$, there are two (closely located) sets of points - squares
denoting data for the square lattice (s)  and hexagons denoting
data for the brick-wall (bw) layer. The lines are least square
fits to the data -- the top two lines correspond to $g=2.0$ and
are fits linear in $1/L$ to the data, while the bottom two lines
correspond to $g=2.2$ and are   polynomials fits of up to order
$1/L^3$ to the data. The data shows that the vanishing of the
structure factor at the critical point occurs simultaneously in
both layers.}
\end{figure}

It is interesting to consider the in-plane magnetization of the
two layers. We note that while for decoupled planes, the
magnetizations of each layer is different, strong interlayer
coupling is seen to make the magnetization uniform in both layers
(see Fig. \ref{Fig4}). For example, for $g=2.0$, while the finite
size values of the in-plane structure factor in the square lattice
is always slightly higher than in the brick-wall layer, the
extrapolated thermodynamic values are the same within the accuracy
of the calculations (see Fig. \ref{Fig4}) -- top two lines). The
in-plane structure factors also lead to conclusions in agreement
with the estimate of the quantum critical point from the total
system magnetization. As can be seen in Fig. \ref{Fig4}, the
N\'{e}el-to-dimer phase transition at $g_c \approx 2.2$ is
associated with the vanishing of the staggered magnetization in
the two layers simultaneously.

\section{Optical lattice implementation \label{sec:oli}}

Realizing these bilayers in solid-state systems, and tuning the
coupling strength between the square and brick-wall layers is
challenging. Fermi gases in optical lattices offer an interesting
alternative where the system parameters can be freely adjusted. In
the following we show that square/brick-wall bilayers can be
generated using an extension of the optical lattice setups already
used to realize brick-wall \cite{leticia-ag} and double-well
lattices \cite{Sebby-Strabley2006, Folling2007,junru16}.

The key ingredient of the experimental implementation that we
propose is a laser setup generating an optical lattice potential
which alternates between square and brick-wall lattice structures
when moving along the $z$ direction. It can be obtained through
the addition of two independent potentials, designed in the
following by $V_{\mathrm{int}}(x,y,z)$ and $V_{\mathrm{sq}}(x,y)$.

\begin{figure}[h]
\centering
\includegraphics[width=0.8\textwidth, angle=0]{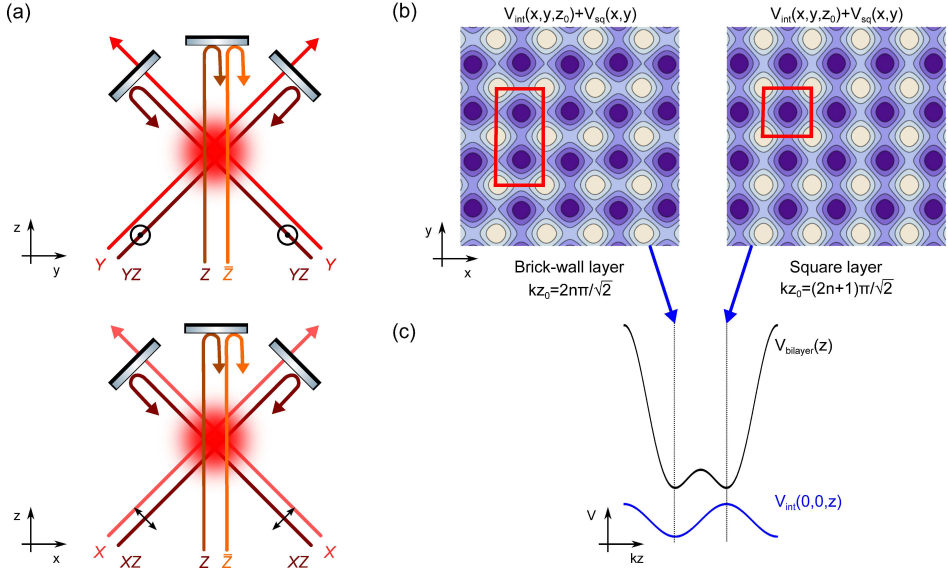}
\caption{\label{Fig8} (a) Beam arrangement for the generation of a
square/brick-wall bilayer optical lattice. Four retroreflected
beams (brown) with out-of-plane and in-plane polarization (black
arrows) interfere in the $xy$ and $xz$ planes respectively,
creating the potential of equation (\ref{eq:Vint}). The addition
of two running wave beams (red) in each plane creates an
additional square lattice structure in the $xy$ plane. A
bichromatic optical lattice potential (orange) is added along the
$z$ direction. (b) Potential landscape in the $xy$ plane for two
different values of $z$, corresponding to brick-wall and square
structures. (c) The bilayer character is obtained by selectively
coupling a square and a brick-wall layer using a double-well
superlattice along the $z$ direction. Its periodicity is adjusted
in order to make its two minima coincide with a maximum and a
minimum of $V_{\mathrm{int}}$, and thus with a brick-wall and a
square plane. The figures correspond to single beam lattice depths
$V_Y/V_X=0.9$, $V_{XZ}/V_{X}=V_{YZ}/V_{X}=0.02$ and
$V_{Z}/V_{\bar{Z}}=2$.}
\end{figure}

$V_{\mathrm{int}}(x,y,z)$ is created by the interference of the
four retro-reflected beams sketched in Fig. \ref{Fig8}(a). Two
retro-reflected beams propagate in the $xz$ plane along $(\pm
1,0,1)/\sqrt{2}$ and have in-plane polarization, whereas two other
beams propagate in the $yz$ plane along $(0,\pm 1,1)/\sqrt{2}$ and
have out-of-plane polarization. Their wavelength $\lambda$ is
chosen to be red-detuned with respect to the atomic transitions.
Hence, the atoms are trapped in the intensity maxima of the light.
When setting the time-phase between the four beams to zero, e.g.
by using standard phase-stabilization schemes (as in Refs.
\cite{Hemerich1992,Greiner2001,leticia-ag,kock16}), the optical
potential reads
\begin{align}
& V_{\mathrm{int}} (x,y,z)
=-V_{XZ}(1+\cos(\sqrt{2}k x)\cos(\sqrt{2}k z))\nonumber\\
&- V_{YZ}(1+ \cos(\sqrt{2}k y)+ \cos(\sqrt{2}k
z)+\cos(\sqrt{2} k y)\cos(\sqrt{2} k z))\nonumber\\
&- 4 \sqrt{2 V_{XZ} V_{YZ}}\cos (k x/\sqrt{2})\cos (k
y/\sqrt{2})\cos^2(k z/\sqrt{2}),\label{eq:Vint}
\end{align}
where $V_{XZ}$ and $V_{YZ}$ denote the single-beam lattice depths
and $k=2 \pi/\lambda$ is the laser wavevector. Superimposing a
standard square lattice potential in the $xy$ plane
\begin{align}
& V_{\mathrm{sq}} (x,y) =-V_{X}\sin^2(k x/\sqrt{2})-V_{Y}\cos^2(k
y/\sqrt{2})\label{eq:Vsquare}
\end{align}
results in the desired succession of brick-wall and square planes
for $k z=2n \pi/\sqrt{2}$ and $k z=(2n+1) \pi/\sqrt{2}$
respectively (where $n$ is an integer index) when choosing
adequate values of the single-beam lattice depths. For instance,
Fig. \ref{Fig8}(b) displays the potential obtained for
$V_Y/V_X=0.9$ and $V_{XZ}/V_{X}=V_{YZ}/V_{X}=0.02$. Note that, in
order to simplify the experimental implementation,
$V_{\mathrm{sq}}(x,y)$ can be realized using the same beam
wavelength and geometry as $V_{\mathrm{int}}(x,y,z)$, but
including only the running-wave beams. Then, a small frequency
detuning $\delta$ between the beams leading to the different
potentials is required in order to avoid unwanted
cross-interference terms. Furthermore, the relative position of
the $V_{\mathrm{int}}$ and $V_{\mathrm{sq}}$ potentials in the
$xy$ plane needs to be adjusted (which can be achieved similarly
to Ref. \cite{Jo2012}).

Finally, to get the desired bilayer structure we add an
independent bichromatic optical lattice along $z$ with lattice
depths $V_Z$ and $V_{\bar{Z}}$, and wavevectors $k_Z$ and
$k_{\bar{Z}}=2 k_Z$
\begin{equation}
V_{\mathrm{bilayer}}(z)=-V_Z\sin^2(k_Z z)-V_{\bar{Z}}\sin^2(2 k_Z
z).
\end{equation}
It selects the $xy$ planes in which the atoms are trapped and
imprints a bilayer pattern along $z$, as shown in Fig.
\ref{Fig8}(c). Tunneling between bilayers is suppressed by a
strong potential barrier, whereas tunneling between two layers of
each bilayer structure can be flexibly controlled by adjusting the
ratio $V_{\bar{Z}}/V_Z$. The wave vectors $k_Z$ and $k$ must be
chosen such that the two potential minima of each bilayer coincide
with a maximum and a minimum of $V_{\mathrm{int}}(0,0,z)$, a
situation which is realized for $k_Z = \sqrt{2} k/3$. As shown in
Fig. \ref{Fig8}(c), the total optical potential
$V(x,y,z)=V_{\mathrm{int}}+V_{\mathrm{sq}}+V_{\mathrm{bilayer}}$
results in a square lattice in one minimum of each bilayer, and a
brick-wall lattice in the other minimum with matching positions of
the potential minima in both layers.

For fermionic $^{6}$Li atoms the lattice potential can be
generated for instance with laser beams at $\lambda_Z=1550$ nm,
$\lambda_{\bar{Z}}=775$ nm and $\lambda=730.7$ nm, which have the
correct wavelength relations and are all red-detuned with respect
to the $D_1$ and $D_2$ transitions at $671$ nm. A polarized gas
can be used to probe the lattice band structure and the properties
of the Dirac points using a combination of Bloch oscillations and
interband transitions as in Ref. \cite{leticia-ag}.
Interactions can then be adjusted to both attractive and repulsive
values using the broad Feshbach resonances available in the three
lowest Zeeman sublevels \cite{Zuern2013}. In the attractive case,
evidence for superfluidity could be obtained by observing
interference peaks in time-of-flight images after sudden release
from the optical lattice \cite{Chin2006}. However, attaining the
required temperatures in the Hubbard regime remains challenging.
The repulsive case is more favorable, and recent experiments have
already demonstrated the emergence of short-range magnetic
correlations in various dimensionalities (although long-range
magnetic order has not yet been observed). The experiments use
either site merging \cite{leticia2, Greif2015}, Bragg scattering
\cite{hulet14}, or high resolution in-situ imaging
\cite{Parsons2016, Boll2016, Cheuk2016, Drewes2016} to directly
measure the singlet fraction, the spin structure factor, and
spin-spin correlation functions. They thus allow for a
comprehensive characterization of the magnetic properties of each
layer, as well as of the singlet correlations expected between
them in the limit of very large interlayer coupling.

\section{Summary} We have considered a bilayer system with combined
brick-wall and square lattice geometry. In this setting which can
be realized with cold atoms in optical lattices, the brick-wall
layer represents artificial graphene with semi-metal band
structure, while the square layer exhibits a normal (metallic)
band structure.

First, we have investigated how the coupling between the layers
modifies the band structure. The Dirac points stemming from the
brick-wall layer are shifted in ${\bf k}$-space, and merge at
$J_\perp/J=2\sqrt{3}$. At larger coupling, a gap is opened, thus the semi-metallic behavior is replaced by band-insulating behavior. 
At criticality, the excitations show exotic anisotropic features: Along one lattice direction, they remain massless, while in the other direction, they are massive.
Additional Dirac points arise from the intersection of brick-wall and square lattice dispersion. Their position in $\bf k$-space is independent from the (finite) strength of the coupling $J_\perp$. 

Second, we have studied superfluidity of the system in the case of
attractive interactions, using a mean-field approach. Similar to a
single brick-wall layer at half filling, we find a semi-metal to
superfluid transition of the bilayer. In the bilayer, such transition not only occurs at filling 1/2, but also at fillings 1/4 and 3/4 when the additional Dirac points appear at the Fermi surface.

Third, we have studied the case of strong repulsive interactions,
in which the half-filled system is in the Mott phase, and can be
mapped onto a Heisenberg antiferromagnet. Using the quantum Monte
Carlo method, we have studied the behavior upon tuning the interlayer
coupling. For $J_\perp/J =2.2$, we find a phase transition
from magnetic long-range order within the planes to a valence
bond crystal with strong dimers between the layers.

Finally, we have presented a possible experimental implementation
which is readily accessible in experiments with Fermi gases in
optical lattices using state-of-the-art technology, and discussed
perspectives for observing experimentally these phases.

Beyond its relevance for artificial graphene systems discussed here,
our approach for generating an atomic bilayer system provides a general
new tool for studying proximity effects with cold atoms. Future work
may combine the bilayer scenario of this paper with ideas from a recent
proposal where parafermions emerge in a heterostructure of a Bose-Einstein
condensate and a fractional quantum Hall system \cite{maghrebi15}.

\ack
We acknowledge support from the European Union
(ERC-2013-AdG Grant No. 339106 OSYRIS, FP7-ICT-2011-9 No. 600645
SIQS, H2020-FETPROACT-2014 No. 641122 QUIC, FP7/2007-2013 Grant
No. 323714 Equam, PCIG13-GA-2013 No. 631633 MagQUPT), from Spanish
MINECO (FIS2013-46768-P FOQUS, FIS2014-59546-P StrongQSIM,
SEV-2015-0522 Severo Ochoa), from the Generalitat de Catalunya
(2014 SGR 874), from German DFG (FOR 2414), from AFOSR-MURI, and from Fundaci\'{o} Cellex.  M.L. thanks the  KITPC Program "Spin-orbit-coupled quantum gases" for support and  hospitality. T.G. acknowledges a JQI Postdoctoral Fellowship.

\vspace{1cm}


\begin{thebibliography}{10}

\bibitem{roadmap-graphene}
K.~S. Novoselov {\it et~al.}, Nature {\bf 490},  192  (2012).

\bibitem{castro_neto}
A.~H. {Castro Neto} {\it et~al.}, Rev. Mod. Phys. {\bf 81},  109  (2009).

\bibitem{bilayer-graphene}
K.~S. Novoselov {\it et~al.}, Nat. Phys. {\bf 2},  177  (2006).

\bibitem{ortix2012}
C. Ortix, L. Yang, and J. van~den Brink, Phys. Rev. B {\bf 86},  081405
  (2012).

\bibitem{yankowitz2012}
M. Yankowitz {\it et~al.}, Nat. Phys. {\bf 8},  382  (2012).

\bibitem{principi12}
A. Principi {\it et~al.}, Phys. Rev. B {\bf 86},  085421  (2012).

\bibitem{gamucci14}
A. Gamucci {\it et~al.}, Nat. Comm. {\bf 5},    (2014).

\bibitem{aliaj16}
I. Aliaj {\it et~al.}, APL Mater. {\bf 4},    (2016).

\bibitem{maciek-ag}
M. Polini {\it et~al.}, Nat. Nano. {\bf 8},  625  (2013).

\bibitem{park2009}
C.-H. Park and S.~G. Louie, Nano Letters {\bf 9},  1793  (2009).

\bibitem{pellegrini}
A. Singha {\it et~al.}, Science {\bf 332},  6034  (2011).

\bibitem{wang16}
{S. Wang \textit{et al.}}, submitted to APL  (2016).

\bibitem{gomesKK}
K.~K. Gomes {\it et~al.}, Nature {\bf 483},  306  (2012).

\bibitem{beenakker07}
R.~A. Sepkhanov, Y.~B. Bazaliy, and C.~W.~J. Beenakker, Phys. Rev. A {\bf 75},
  063813  (2007).

\bibitem{RechtsmanPRL2013}
M.~C. Rechtsman {\it et~al.}, Phys. Rev. Lett. {\bf 111},  103901  (2013).

\bibitem{KuhlPRB2010}
U. Kuhl {\it et~al.}, Phys. Rev. B {\bf 82},  094308  (2010).

\bibitem{soltan-hexa}
P. Soltan-Panahi {\it et~al.}, Nat. Phys. {\bf 7},  434  (2011).

\bibitem{leticia-ag}
L. Tarruell {\it et~al.}, Nature {\bf 483},  302  (2012).

\bibitem{mlbook}
{M. Lewenstein}, {A. Sanpera}, and V. Ahufinger, {\em {Ultracold Atoms in
  Optical Lattices - Simulating quantum many-body systems}} (Oxford University
  Press, New York, 2012).

\bibitem{Jo2012}
G.-B. {Jo} {\it et~al.}, Phys. Rev. Lett. {\bf 108},  045305  (2012).

\bibitem{DucaScience2015}
L. Duca {\it et~al.}, Science {\bf 347},  288  (2015).

\bibitem{LiScience2016}
T. {Li} {\it et~al.}, Science {\bf 352},  1094  (2016).

\bibitem{weinberg2016}
M. Weinberg {\it et~al.}, 2D Materials {\bf 3},  024005  (2016).

\bibitem{UehlingerEPJB2013}
T. {Uehlinger} {\it et~al.}, EPJ ST {\bf 217},  121  (2013).

\bibitem{uehlinger}
T. Uehlinger {\it et~al.}, Phys. Rev. Lett. {\bf 111},  185307  (2013).

\bibitem{jotzu14}
G. Jotzu {\it et~al.}, Nature {\bf 515},  237  (2014).

\bibitem{FlaeschnerScience2016}
N. {Fl{\"a}schner} {\it et~al.}, Science {\bf 352},  1091  (2016).

\bibitem{duan-ag}
S.-L. Zhu, B. Wang, and L.-M. Duan, Phys. Rev. Lett. {\bf 98},  260402  (2007).

\bibitem{sols}
B. Wunsch, F. Guinea, and F. Sols, New J. Phys. {\bf 10},  103027  (2008).

\bibitem{LeePRA2009}
K.~L. Lee {\it et~al.}, Phys. Rev. A {\bf 80},  043411  (2009).

\bibitem{poletti2011}
D. Poletti, C. Miniatura, and B. Gr{\'e}maud, EPL {\bf 93},  37008  (2011).

\bibitem{luehmann2014}
D.-S. L{\"u}hmann {\it et~al.}, Phys. Rev. A {\bf 90},  013614  (2014).

\bibitem{cao2015}
L. Cao {\it et~al.}, Phys. Rev. A {\bf 91},  043639  (2015).

\bibitem{lutchyn2010}
R.~M. Lutchyn, J.~D. Sau, and S. {Das Sarma}, Phys. Rev. Lett. {\bf 105},
  077001  (2010).

\bibitem{sau2010}
J.~D. Sau, R.~M. Lutchyn, S. Tewari, and S. {Das Sarma}, Phys. Rev. Lett. {\bf
  104},  040502  (2010).

\bibitem{hui2015}
H.-Y. Hui, J.~D. Sau, and S. {Das Sarma}, Phys. Rev. B {\bf 92},  174512
  (2015).

\bibitem{cole2016}
{W. S. Cole}, {J. D. Sau}, and S. {Das Sarma}, arXiv/  1603.03780  (2016).

\bibitem{wind15}
{D. Scarabelli \textit{et al.}}, arxiv/1507.04390  (2015).

\bibitem{montambaux09}
G. Montambaux, F. Pi{\'e}chon, J.-N. Fuchs, and M.~O. Goerbig, Phys. Rev. B
  {\bf 80},  153412  (2009).

\bibitem{dietl08}
P. Dietl, F. Pi{\'e}chon, and G. Montambaux, Phys. Rev. Lett. {\bf 100},
  236405  (2008).

\bibitem{paramekanti06}
E. Zhao and A. Paramekanti, Phys. Rev. Lett. {\bf 97},  230404  (2006).

\bibitem{paramekanti12}
S. Tsuchiya, R. Ganesh, and A. Paramekanti, Phys. Rev. A {\bf 86},  033604
  (2012).

\bibitem{PhysRevB.23.1447}
S. Robaszkiewicz, R. Micnas, and K.~A. Chao, Phys. Rev. B {\bf 23},  1447
  (1981).

\bibitem{PhysRevB.24.1579}
S. Robaszkiewicz, R. Micnas, and K.~A. Chao, Phys. Rev. B {\bf 24},  1579
  (1981).

\bibitem{PhysRevB.24.4018}
S. Robaszkiewicz, R. Micnas, and K.~A. Chao, Phys. Rev. B {\bf 24},  4018
  (1981).

\bibitem{PhysRevLett.66.946}
A. Moreo and D.~J. Scalapino, Phys. Rev. Lett. {\bf 66},  946  (1991).

\bibitem{ho09}
A.~F. Ho, M.~A. Cazalilla, and T. Giamarchi, Phys. Rev. A {\bf 79},  033620
  (2009).

\bibitem{PhysRevB.47.7995}
D.~J. Scalapino, S.~R. White, and S. Zhang, Phys. Rev. B {\bf 47},  7995
  (1993).

\bibitem{PhysRevB.80.245118}
K.~L. Lee {\it et~al.}, Phys. Rev. B {\bf 80},  245118  (2009).

\bibitem{review-heisenberg-square}
E. Manousakis, Rev. Mod. Phys. {\bf 63},  1  (1991).

\bibitem{leticia2}
D. Greif {\it et~al.}, Science {\bf 340},  1307  (2013).

\bibitem{hulet14}
R.~A. Hart {\it et~al.}, Nature {\bf 519},  211  (2015).

\bibitem{Greif2015}
D. {Greif} {\it et~al.}, Phys. Rev. Lett. {\bf 115},  260401  (2015).

\bibitem{Parsons2016}
M.~F. Parsons {\it et~al.}, Science {\bf 353},  1253  (2016).

\bibitem{Boll2016}
M. Boll {\it et~al.}, Science {\bf 353},  1257  (2016).

\bibitem{Cheuk2016}
L.~W. Cheuk {\it et~al.}, Science {\bf 353},  1260  (2016).

\bibitem{Drewes2016}
J.~H. {Drewes} {\it et~al.}, arXiv/1607.00392  (2016).

\bibitem{Sandvik0}
A.~W. Sandvik, Phys. Rev. B {\bf 56},  11678  (1997).

\bibitem{Castro}
E.~V. Castro, N.~M.~R. Peres, K.~S.~D. Beach, and A.~W. Sandvik, Phys. Rev. B
  {\bf 73},  054422  (2006).

\bibitem{Sandvik-Scalapino}
A.~W. Sandvik and D.~J. Scalapino, Phys. Rev. Lett. {\bf 72},  2777  (1994).

\bibitem{Ganesh}
R. Ganesh, S.~V. Isakov, and A. Paramekanti, Phys. Rev. B {\bf 84},  214412
  (2011).

\bibitem{HighTcAFM}
S. Sachdev and J. Ye, Phys. Rev. Lett. {\bf 69},  2411  (1992).

\bibitem{Chubukov}
A.~V. Chubukov and S. Sachdev, Phys. Rev. Lett. {\bf 71},  169  (1993).

\bibitem{Sokol}
A. Sokol and D. Pines, Phys. Rev. Lett. {\bf 71},  2813  (1993).

\bibitem{bec}
T. Nikuni, M. Oshikawa, A. Oosawa, and H. Tanaka, Phys. Rev. Lett. {\bf 84},
  5868  (2000).

\bibitem{GiamarchiNaturePhys2008}
T. {Giamarchi}, C. {R{\"u}egg}, and O. {Tchernyshyov}, Nat. Phys. {\bf 4},  198
   (2008).

\bibitem{Wang}
L. Wang, K.~S.~D. Beach, and A.~W. Sandvik, Phys. Rev. B {\bf 73},  014431
  (2006).

\bibitem{Sandvik-vbs}
A.~W. Sandvik, Phys. Rev. Lett. {\bf 95},  207203  (2005).

\bibitem{Lin}
Y.-J. Lin {\it et~al.}, Nature {\bf 462},  628  (2009).

\bibitem{Huse}
D.~A. Huse, Phys. Rev. B {\bf 37},  2380  (1988).

\bibitem{Sebby-Strabley2006}
J. {Sebby-Strabley}, M. {Anderlini}, P.~S. {Jessen}, and J.~V. {Porto}, Phys.
  Rev. A {\bf 73},  033605  (2006).

\bibitem{Folling2007}
S. {F{\"o}lling} {\it et~al.}, Nature {\bf 448},  1029  (2007).

\bibitem{junru16}
J. Li {\it et~al.}, Phys. Rev. Lett. {\bf 117},  185301  (2016).

\bibitem{Hemerich1992}
A. Hemmerich, D. Schropp, T. Esslinger, and T.~W. H{\"a}nsch, Eur. Phys. Lett.
  {\bf 18},  391  (1992).

\bibitem{Greiner2001}
M. {Greiner} {\it et~al.}, Phys. Rev. Lett. {\bf 87},  160405  (2001).

\bibitem{kock16}
T. Kock, C. Hippler, A. Ewerbeck, and A. Hemmerich, J. Phys. B {\bf 49},
  042001  (2016).

\bibitem{Zuern2013}
G. {Z{\"u}rn} {\it et~al.}, Phys. Rev. Lett. {\bf 110},  135301  (2013).

\bibitem{Chin2006}
J.~K. {Chin} {\it et~al.}, Nature {\bf 443},  961  (2006).

\bibitem{maghrebi15}
M.~F. Maghrebi {\it et~al.}, Phys. Rev. Lett. {\bf 115},  065301  (2015).

\end{thebibliography}

\end{document}